\documentclass[twocolumn, secnumarabic, amssymb, amsmath,nobibnotes, aps, prl,nofootinbib]{revtex4}
\begin{document}
\title{The Thermodynamics of the Hagedorn Mass Spectrum}
\author{B. H. Lavenda}
\email{bernard.lavenda@unicam.it}
\affiliation{Universit$\grave{a}$ degli Studi, Camerino 62032 (MC) Italy}
\date{\today}
\newcommand{\sumn}{\sum_{i=1}^{n}\,}
\newcommand{\sumk}{\sum_{i=1}^{k}\,}
\newcommand{\half}{\mbox{\small{$\frac{1}{2}$}}}
\newcommand{\fourth}{\mbox{\small{$\frac{1}{4}$}}}\newcommand{\twothirds}{\mbox{\small{$\frac{2}{3}$}}}
\newcommand{\nn}{\mbox{\small{$\frac{1}{n}$}}}
\newcommand{\Th}{T_{\mbox{\tiny{H}}}}
\newcommand{\En}{E_{\mbox{\tiny{N}}}}
\newcommand{\Vn}{V_{\mbox{\tiny{N}}}}
\newcommand{\Sn}{S_{\mbox{\tiny{N}}}}
\newcommand{\betah}{\beta_{\mbox{\tiny{H}}}}
\newcommand{\Fn}{F_{\mbox{\tiny{N}}}}
\newcommand{\pf}{p_{\mbox{\tiny{$\infty$}}}}
\newcommand{\Ef}{E_{\mbox{\tiny{$\infty$}}}}
\newcommand{\Nf}{N_{\mbox{\tiny{$\infty$}}}}

\begin{abstract}
No bootstrap assumption is needed  to derive the exponential growth of the Hagedorn hadron mass spectrum: It is a consequence of the second law applied to a relativistic gas, and the relativistic equivalence between inertial mass and its heat content.  The Hagedorn temperature occurs in the limit as  the number of particles and their internal energy diverge such that their ratio remains constant. The divergences in the $N$ particle entropy, energy, and free energy result when this condition is imposed upon a mixture of ideal gases, one conserving particle number and the other not. The analogy with a droplet in the presence of vapor  explains why the pressure of the droplet continues to increase as the temperature rises finally leading to its break up when the Hagedorn temperature is reached. The adiabatic condition relating the particle volume to the Hagedorn temperature is asymptotic.  Since it is a limiting temperature, and not a critical one, there can be no phase transition of whatever kind, and the original density of states used to derive such a phase transition is not thermodynamically admissible because its partition function does not exist.
\end{abstract}
 
\maketitle
\section{The second law and the hagedorn mass spectrum}
Hagedorn's hypothesis of an exponential growth in the number of hadronic resonances has become a central pillar to particle physics \cite{Bron}. His formula for the asymptotic dependence of the density of hadronic states upon mass is \cite{Hag}
\begin{equation}
\Omega(m)\approx\exp\left(m/T_{\mbox{\tiny H}}\right), \label{eq:Hag}
\end{equation}
apart from a prefactor which is a \lq slowly varying\rq\ function \cite{Bron} of the mass $m$ of the hadron, and $T_{\mbox{\tiny H}}$ is the Hagedorn temperature associated with its mass spectrum. This temperature is supposedly the highest attainable temperature.\par In this letter, we show that  (i) the exponential mass growth in (\ref{eq:Hag}) is due to the second law applied to a perfect relativistic  gas (prg) (e.g. a pion gas), and (ii) the existence of a limiting temperature is a consequence of the application of Cocconi's assumption \cite{Coc} that the ratio of the total energy to the particle number should tend to a constant as the energy and number of particles increase without limit to a mixture of non-interacting prg and pcg, which behave as a droplet in equilibrium with the surrounding vapor. The eventual break up of the droplet occurs in the Hagedorn limit where the pressure diverges.\par Consequently, the second law applied to a prg and the relativistic equivalence of inertial mass and heat makes  the \lq bootstrap\rq\ assumption  superfluous to the derivation of the hadron mass spectrum. Moreover, the form of the prefactor in (\ref{eq:Hag}) is constrained by the existence of its Laplace transform.\par
Let $T$ and $S$ stand for the absolute temperature and metrical entropy, respectively. The distinction between the metrical, $S$, and empirical, $\sigma$, entropies is that whereas the former is defined up to an arbitrary increasing linear continuous transformation, the latter is defined up to an arbitrary, continuous, and strictly increasing scale transformation. Assuming that $T$ is some increasing function of the empirical temperature, we have
\begin{equation}
T(t)\,dS(\sigma)=T(t)S^{\prime}(\sigma)\,d\sigma. \label{eq:II}
\end{equation}
To proceed further we need explicit forms for the metrical entropy.\par
From the exactness condition
\begin{equation}
T^{\prime}(t)S^{\prime}(\sigma)=\frac{\partial(V,p)}{\partial(\sigma,t)}=:\mathcal{J} \label{eq:J}
\end{equation}
for the internal energy,
\[dE=T(t)\,dS(\sigma)-p\,dV,\]
it follows that the Jacobian, $\mathcal{J}$ is always the product of two functions, one depending solely on $\sigma$, and the other solely on $t$ \cite{Cara}.\par
A pcg has isotherms
\begin{equation}
pV=t^r \label{eq:pV}
\end{equation}
and adiabats
\begin{equation}
\sigma=V^st^r=pV^{1+s}, \label{eq:sigma}
\end{equation}
where $r\ge1$ so that the empirical temperature cannot increase at a slower rate than the absolute temperature. The exponent, $1/s>1$, will be related to half the degrees of freedom.\par Since 
\begin{equation}
\mathcal{J}_{\mbox{\tiny{pcg}}}=\frac{rt^{r-1}}{s\sigma}, \label{eq:J-c}
\end{equation}
is the Jacobian of the pcg,  the absolute temperature and metrical entropy are
\begin{equation}
T(t)=C^{-1}t^r, \label{eq:T}
\end{equation}
and 
\begin{equation}
S(\sigma)=C\ln\sigma^{1/s}, \label{eq:S-c}
\end{equation}
respectively, where $C>0$ is an arbitrary constant, which we shall set equal to unity in the following.\par
Now consider a prg, whose hallmark is a pressure that is independent of the volume, like that exhibited in a first-order phase transition, or a photon or pion gas. The volume expands, or contracts, to accommodate an influx, or efflux, of particles from one \lq phase\rq\ to another so as to keep the pressure and temperature  constant. The \lq phases\rq\ simply refer to the creation and annihilation of particles so that    the absolute temperature will increase more slowly than the energy, eventually tending to a constant as the energy and particle number grow without limit \cite{Coc}.\par In place of (\ref{eq:pV}) we have
\begin{equation}
p=st^q, \label{eq:p}
\end{equation}
for some positive exponent $q>r$. Whereas the exponent $q$ is related to the total number of degrees of freedom, the exponent $r$ refers to those degrees of freedom that affect the average kinetic energy (i.e. temperature) of the gas.\par The Jacobian for a prg is
\begin{equation}
\mathcal{J}_{\mbox{\tiny prg}}=\frac{q}{r}\sigma^{1/s-1}\,rt^{q-r/s-1}. \label{eq:J-u}
\end{equation}
The absolute temperature should still be proportional to a positive power of the empirical temperature given by (\ref{eq:T}), and this imposes the condition $s=r/(q-r)$. $s$ appears in the exponent $\varepsilon^{1/s-1}\,d\varepsilon$ for the number of states between energy densities, $\varepsilon$ and $\varepsilon+d\varepsilon$, which in turn implies
\begin{equation}
sE=pV, \label{eq:sE}
\end{equation}
from the grand canonical partition function, even when the energy cannot be written as $E=\varepsilon V$ (i.e. for a pcg).\footnote{Expression (\ref{eq:sE}) also follows from the adiabatic condition 
\[
EV^s=\mbox{const.}, \] by differentiating it with respect to $V$, and using the definition of pressure, $p=-\left(\partial E/\partial V\right)_S$.}
\par
From (\ref{eq:J-u}) it follows that
\[S^{\prime}(\sigma)=\frac{q}{r}\sigma^{1/s-1},\]
which upon integration gives
\begin{equation}
S(\sigma)=\frac{q}{q-r}\sigma^{1/s}. \label{eq:S-r}
\end{equation}
Hence, \emph{in contrast to the metrical entropy of a pcg, (\ref{eq:S-c}), the entropy of a prg, (\ref{eq:S-r}), is its exponential\/}. Herein lies the origin of the Hagedorn mass spectrum.
\section{Thermodynamics of high energy particle physics}
The string of inequalities can now be completed in (\ref{eq:II}). For a pcg we have
\[
T(t)S^{\prime}(\sigma)\,d\sigma=\frac{t^r}{s\sigma}\,d\sigma=\frac{d\sigma}{sV^s}=dQ, \]

while for a prg,
\begin{equation}
T(t)S^{\prime}(\sigma)\,d\sigma=t^r\,\frac{q}{r}\sigma^{1/s-1}\,d\sigma=\frac{d\sigma^{q/r}}{V^s}=dQ, \label{eq:II-u}
\end{equation}
where we have used  the definition of an adiabat, (\ref{eq:sigma}).
\par
 In fact,  a new adiabatic potential \cite{Lav}, 
\begin{equation}
L(\sigma)=\sigma^{q/r}, \label{eq:L}
\end{equation} can be defined from (\ref{eq:II-u}),
$dL(\sigma)=V^s\,dQ$,
 indicating that $V^s$, like $1/T$, is a integrating factor for the heat. The reason is that their ratio is the empirical entropy, (\ref{eq:sigma}). However, unlike the entropy, (\ref{eq:S-c}), (\ref{eq:L}) is a not a first-order homogeneous function, and this fact makes the two adiabatic potentials, $L$ and $S$, comparable for deformations as well as thermal interactions, unlike $E$ and $S$, which are both first-order homogeneous functions \cite{Lav}.\par
 According to the definition of metrical entropy, if we allow two systems $A$ and $B$ to interact thermally that are at the same temperature, the composite system $C$ will be given by \cite{Buch}
\[S_C^{\prime}(\sigma_A\sigma_B)\,d\sigma_C(\sigma_A\sigma_B)=S_A^{\prime}(\sigma_A)\,d\sigma_A+S_B^{\prime}(\sigma_B)\,d\sigma_B.\]
For a pcg, the empirical entropy of the composite system is $\sigma_C=\sqrt{\sigma_A\sigma_B}$, so that the metrical entropy of the composite system is 
\begin{equation}
S_C(\sigma_C)=2\ln\sigma_C=\ln\left(\sigma_A\sigma_B\right)=\ln\sigma_A+\ln\sigma_B.\label{eq:geo}
\end{equation}
\par
In contrast, for a prg the metrical entropy is given by the power law, (\ref{eq:S-r}), so that the entropy of the composite system is 
\begin{equation}
S_C(\sigma_C)=2\frac{q}{q-r}\sigma_C^{1/s}=\frac{q}{q-r}\left(\sigma_A^{1/s}+\sigma_B^{1/s}\right),\label{eq:pow}
\end{equation}
where 
\[\sigma_C=\left(\frac{\sigma_A^{1/s}+\sigma_B^{1/s}}{2}\right)^s,\]
is the power mean of order $1/s$.\par The important difference between (\ref{eq:geo}) and (\ref{eq:pow}) is that only the latter permits the definition of a norm \cite{Lav}, \begin{equation}\parallel\sigma\parallel\;\; :=2^s\sigma_C=\left(\sigma_A^{1/s}+\sigma_B^{1/s}\right)^{s},\label{eq:norm}
\end{equation} since a norm requires $s\le1$. It is easy to verify that the (\ref{eq:norm}) satisfies the triangle inequality, \[\parallel\sigma_A+\sigma_B\parallel\;\;\le\;\;\parallel\sigma_A\parallel+\parallel\sigma_B\parallel.\] It also explains why conventional thermodynamics, with power density of states and geometric means, cannot define distances.\par
According to the  principle  of latent, $M$, and specific, $N$, heats, the heat required to alter the volume from $V$ to $V+dV$, and the temperature from $T$ to $T+dT$, is
\begin{equation}dQ=M\,dV+N\,dT.\label{eq:Q}
\end{equation}
The second laws, obtained by using the integrating factors $1/T$ and $V^s$, result in the Clapeyron equations \cite{Lav}
\begin{equation}
\left(\frac{\partial p}{\partial T}\right)_V=\frac{M}{T}=\frac{sN}{V}. \label{eq:Clap}
\end{equation}
\par
It has been claimed that empirical entropy shares with empirical temperature the property they are defined up to scale factor \cite[pp. 70-71]{Buch},
\[dQ=\tau\,d\sigma=\tau^{\star}\,d\sigma^{\star},\]
where $\tau$ and $\tau^{\star}$ are integrating denominators for $dQ$. If we take the scaling $\sigma^{\star}=\sigma^{1/s}$, then 
\begin{equation}
\tau^{\star}=\tau\bigg/(1/s)\sigma^{1/s-1}\label{eq:tau}
\end{equation}
is an integrating denominator for $dQ$.\par Even more can be said: Any function of $\sigma$ that multiplies the integrating factor will satisfy the exactness condition for a perfect differential because the condition,
\[M\left(\frac{\partial\ln\sigma}{\partial T}\right)_V=N\left(\frac{\partial\ln\sigma}{\partial V}\right)_T,\]
is satisfied on the strength of (\ref{eq:Clap}), and the definition of the empirical entropy. 
\par
\section{Classical and relativistic joule-thomson effects}
Enthalpy plays a very special role for both  a pcg and prg. It is well known that the enthalpy is conserved in the Joule-Thomson effect. To see how conservation arises, we write (\ref{eq:Q}) in the form
\begin{equation}
dQ=p\,dV+\frac{V}{s}\left(\frac{\partial p}{\partial T}\right)_V dT. \label{eq:Q-c}
\end{equation}
Introducing the total differential $dp$ in the second term of (\ref{eq:Q-c}) we get
\begin{equation}
dQ=p\,dV+\frac{1}{s}\,d(pV).\label{eq:Q-c-bis}
\end{equation}
\par
When integrated between the two volumes on both sides of the porous plug, the first term on the right-hand side is the work that is converted entirely into heat by friction, while the integrated form of the second term is the total change in internal energy due to the total work expended on the gas to drive it through the porous plug. Therefore,  heat,
\begin{equation}
Q=\int_{V}^{V^{\prime}}\,p\,dV+E^{\prime}-E,\label{eq:I}
\end{equation}
is absorbed as the gas expands from volume $V$ to volume $V^{\prime}$. Since  the pressure varies as the gas is driven through the plug, we add the integral of $V\,dp$ to both sides of (\ref{eq:I}) to get
\begin{equation}H^{\prime}-H=p^{\prime}V^{\prime}-pV+E^{\prime}-E.\label{eq:H}
\end{equation}
\par
At this point, the adiabatic condition is invoked to get the isoenthalpic condition, $H^{\prime}=H$. However, the first law is (\ref{eq:I}), and it has been derived from (\ref{eq:Q-c-bis}) in which the heat transfer does not vanish. So if (\ref{eq:H}) is to vanish it must be a separate conservation unrelated to the condition of adiabaticity.\par
The Joule-Thomson effect for a prg is equally as important. Since the pressure is independent of the volume, we get immediately
\begin{equation}
dQ=\left(1+\frac{1}{s}\right)p\,dV+\frac{1}{s}V\,dp, \label{eq:Q-r}
\end{equation}
for the heat absorbed, where we have introduced the latent heat $M=\varepsilon+p$, and $\varepsilon=p/s$.  Adding $V\,dp$ to both sides makes the right-hand side a total differential, whose integral is
\begin{equation}
H=\frac{q}{r}pV=TS, \label{eq:H-r}
\end{equation}
where the last equality follows from $pV+E=TS$, since the chemical potential of a prg vanishes identically \cite{Lav}.\par Planck \cite{Plan} was the first to appreciate that the enthalpy was proportional to the momentum.  He referred to
\begin{equation}
\rho=\frac{H}{V}, \label{eq:Planck}
\end{equation}
as the \lq law of inertia of energy\rq\ in units where $c=\kappa \;\;(\mbox{Boltzmann's constant}) =1$. That is, the forces acting on a body are transmitted by a momentum density whose source is a flow of energy.\par Now, (\ref{eq:Q-r}) can be written as
\begin{equation}
dQ=h\,dV+V\,d\varepsilon, \label{eq:Q-r-bis}
\end{equation}
and transformed into
\[dQ=d(hV)-V\,dp=T\,d(sV),\]
by using the thermodynamic identity, $d(hV)-V\,dp=T\,d(sV)$, where $h$ and $s$ are the enthalpy and entropy densities, respectively. Hence, the process is not adiabatic.\par In fact, (\ref{eq:Q-r}) can be converted into an energy balance relation. Equation (\ref{eq:Q-r-bis}) is equivalent to
\begin{equation}{\dot{Q}}=h\,{\dot{V}}+V\,{\dot{\varepsilon}},\label{eq:Q-dot}
\end{equation}
where the dot denotes the substantial derivative. Using the equation of continuity, ${\dot{V}}=V\nabla\cdot u$, where $1/V$ is the density, and $u$ the velocity vector, (\ref{eq:Q-dot}) can be written as the energy balance equation
\begin{equation}
\frac{\partial\varepsilon}{\partial t}+\nabla\cdot J=q-u\cdot\nabla p, \label{eq:E-con}
\end{equation}
or using the balance equation for the kinetic energy, 
\[\frac{\partial\half u^2/V}{\partial t}+\nabla\cdot\left(\half u^2/V\right)=-u\cdot\nabla p,\]
it can be converted  into the total energy balance equation,
\[\frac{\partial\left(\varepsilon+\half u^2/V\right)}{\partial t}+\nabla\cdot u\left(h+\half u^2/V\right)=q,\]
where $J=hu$ is the Poynting vector in (\ref{eq:E-con}), and $q=\dot{Q}/V$, the rate of Joule heating per unit volume.
\par
Thus, the relativistic Joule-Thomson effect, like its nonrelativistic analog, is isenthalpic, but not adiabatic. In order for the two to coincide the pressure must be maintained constant throughout the process, which it is not, for, otherwise, we could have integrated the integral in (\ref{eq:I}) without further ado.\par \emph{\lq Heating\rq\ is a form of energy flux, and, relativistically, the energy flux is proportional to the momentum flux\/}. This is the origin of the mass-enthalpy density relation (\ref{eq:Planck}).\par We will now  show that the relativistic equivalence of inertial mass and heat, (\ref{eq:Planck}), and the relation of the entropy to the density of states, (\ref{eq:S-r}), will  yield the Hagedorn mass spectrum (\ref{eq:Hag}).\par
\section{Derivation of the Hagedorn mass spectrum}
For a pcg, the structure function is
\[
\Omega(\sigma)=\sigma^{1/s}, \]
while for a prg, we have
\begin{equation}
\Omega(\sigma)=\exp\left(\frac{q}{q-r}\sigma^{1/s}\right). \label{eq:Omega-r}
\end{equation}
No matter which form applies, the entropy per particle is 
\begin{equation}
S(\sigma)=\ln\Omega(\sigma), \label{eq:Boltz}
\end{equation}
which is  Boltzmann's principle.\par
Introducing (\ref{eq:H-r}) into (\ref{eq:Omega-r}), and requiring the equivalence of internal mass and heat, result in
\[
\Omega(m)=\exp\left(\frac{m}{T}\right), \]
where $m=\rho V$, the mass in the particle volume $V$ whose density is $\rho$.
 \emph{The exponential increase in the density of states with mass is the result of the fact that the pressure is independent of the volume, and is a sole function of the temperature.\/} Classically, this corresponds to a two phase system as in a first-order phase transition. We have, as yet, to explain the existence of a limiting temperature.
\par 
Suppose that both pcg and prg are present and in equilibrium with each other [cf. (\ref{eq:equi}) below]. The entropy per particle is then  the sum of (\ref{eq:S-c}) and (\ref{eq:S-r}), viz.
\begin{equation}
S(T,V)=\ln\sigma_c^{1/s}+\frac{q}{q-r}\sigma_r^{1/s},\label{eq:S+S}
\end{equation}
where the subscripts $c$ and $r$ refer to a pcg and a prg, respectively. Whereas in the former, the temperature is $T=sE$, the temperature of the latter is $T=\left(E/V\right)^{r/q}$, showing that the temperature of a prg increases more slowly than its energy, $r<q$. Thus, in terms of the energy, (\ref{eq:S+S}) is
\begin{equation}
S(E,V)=\ln\left(E^{1/s}V\right)+\frac{q}{q-r}E^{(q-r)/q}V^{r/q}. \label{eq:S+S-bis}
\end{equation}
\par
The idea behind the existence of limiting temperature lies in a statement made by Cocconi \cite{Coc} to the effect that for increasing energy, the number of particles and their kind increases so as to keep the energy particle, and hence the temperature, constant
\begin{equation}
sE=\Th. \label{eq:Coc}
\end{equation}
\par
The temperature of the gas mixture follows from the second law applied to (\ref{eq:S+S-bis}), viz.
\begin{equation}
\left(\frac{\partial S}{\partial E}\right)_V=\frac{1}{sE}+\left(\frac{V}{E}\right)^{r/q}=\frac{1}{T}. \label{eq:1/T}
\end{equation}
Invoking Cocconi's condition (\ref{eq:Coc}) results in
\begin{equation}
\En=s^{-1}N(\Vn,T) \Th=\epsilon^{-q/r}\Ef, \label{eq:En}
\end{equation}
where $\En$ and $\Vn$ are the total energy and volume of 
\begin{equation}
N\left(\Vn,T\right)=\epsilon^{-q/r}\Nf\label{eq:N}
\end{equation}
particles.\par In expressions (\ref{eq:En}) and (\ref{eq:N}) we have introduced the reduced temperature $\epsilon :=(\Th-T)/\Th$, and $\Ef=\Vn T^{q/r}$ and $\Nf=s\Vn T^{1/s}$ are the energy and total number of particles of the prg in the absense of any constraint upon the temperature. The effect of (\ref{eq:Coc}) is to bring the number of particles to the same power of the reduced temperature as the energy.\par  In the limit $\epsilon\rightarrow1$, (\ref{eq:En}) reduces to the generalized Stefan-Boltzmann law for a prg, whereas, in the limit $\epsilon\rightarrow0$, the total energy and number of particles, become infinite.\par
This is confirmed by the definition of the pressure,
\[\left(\frac{\partial S}{\partial V}\right)_E=\frac{1}{V}+s\left(\frac{E}{V}\right)^{(q-r)/q}=\frac{p}{T}.\]
The presence of the first term implies implicitly the existence of an upper limit to the work that the gas can do. For upon introducing (\ref{eq:Coc}) there results
\[\En=s^{-1}N(\Vn,T)\Th=\Vn\left(\frac{T\Th}{pV-T}\right)^{q/r},\]
which will coincide with (\ref{eq:En}) if $pV=\Th$. This together  with (\ref{eq:Coc}) gives the equation of state (\ref{eq:sE}).\par
Introducing the logarithm of the partition function, $Z$, by
\begin{equation}
-\left(\frac{\partial}{\partial\beta}\ln Z(\beta,V)\right)_V=E=V\left(\beta-\betah\right)^{-q/r},\label{eq:EE}
\end{equation}
where $\beta=1/T$, and integrating we get
\begin{equation}
\ln Z(\beta,V)=s V\left(\beta-\betah\right)^{-1/s}=-\beta F,\label{eq:Z}
\end{equation}
for $q\neq r$, while
\[
\ln Z(\beta,V)=-V\ln\left(\beta-\betah\right)=-\beta F, \]
for $q=r$.  $F$ is the  Helmholtz free energy per particle.\par Apart from constant factors, the \emph{logarithm\/} of the partition function (\ref{eq:Z}) is identical to Hagedorn's partition function \cite[formula (31)]{Hag}, with $1/s$ replacing his exponent $\alpha$ in the density of states. This means that the density of states would be given essentially by a power law for a pcg, and not an exponential as for a prg [cf. (\ref{eq:dos}) below].\par
The density of states can now be obtained by an inverse Laplace transform on the partition function, $Z$, or, equivalently, in the high temperature limit, to a Legendre transform of (\ref{eq:Z}). This gives the entropy per particle as
\begin{eqnarray*}
S(E,V) & = & \frac{E}{\Th}+\frac{q}{q-r}E^{(q-r)/q}V^{r/q}\\
& = & \frac{E}{q-r}\left(\frac{q}{T}-\frac{r}{\Th}\right),
\end{eqnarray*}
for $q\neq r$, where  (\ref{eq:1/T}) has been introduced in the second line, and
\[
S(E,V)=V\left(\ln\frac{E}{V}+1\right)+\frac{E}{\Th}, \]
for $q=r$.\par
The phase equilibrium between the pcg and the prg resembles that of a droplet in contact with vapor. The surface of the droplet acts as a semipermeable membrane which allows vapor to pass but not liquid. As a consequence, the pressure, $p$, of the droplet will be higher than the surrounding pressure of the vapor, $\pf$, according to
\begin{equation}p=\epsilon^{-q/r}\pf.\label{eq:p-drop}
\end{equation}
For $\epsilon\simeq 1$, we have
\begin{equation}
\Delta p:=2\frac{\alpha}{a}=p-\pf\simeq\frac{q}{r}\frac{T}{\Th}\pf, \label{eq:Delta}
\end{equation}
where $\alpha$ is the surface tension, and $a$ is the radius of the droplet.\par
The condition for equilibrium between droplet and vapor  is
\begin{equation}
V^{\prime}dp^{\prime}=V\,dp,\label{eq:equi}
\end{equation}
where the prime denotes the droplet. If we treat the liquid in the droplet as a prg and the vapor as a pcg, (\ref{eq:equi}) can be integrated between the limits
\[
V^{\prime}\int_{\pf}^{\pf+\Delta p}\,dp^{\prime}=T\int_{\pf}^{p}\,\frac{dp}{p}.\]
Two fundamental properties of the gases have been used: For a prg the pressure is independent of volume, while for a pcg the pressure is given by the ideal gas law, $p=T/V$. Performing the integration, we get the well-known result
\[\ln\frac{p}{\pf}=\frac{2\alpha V^{\prime}}{aT}.\] 
\par
Not only can it be concluded that the pressure of the droplet increases as it size decreases, but, moreover, we have from (\ref{eq:Delta}) that
\begin{equation}
V^{\prime}\,\Delta p=\frac{T}{\Th}\,Q,\label{eq:sur}
\end{equation}
where the heat $Q=MV^{\prime}=(q/r)\pf V^{\prime}$, coming from a heat bath at temperature $\Th$, is absorbed by the droplet at temperature $T$ to increase its vapor pressure by an amount $\Delta p$. According to (\ref{eq:Delta}), the left-hand side of (\ref{eq:sur}) is the derivative of the surface free energy, $4\pi a^2\alpha$, of the drop with respect to the number of particles, $4\pi a^3/3V^{\prime}$, of  liquid in the drop, viz. $2\alpha V^{\prime}/a$.
\par
As the liquid evaporates  the droplet becomes smaller, its pressure  increases, initially exponentially, 
\[p=\pf\exp\left(Q/\Th\right), \]and, ultimately,  (\ref{eq:p-drop}) predicts the   break up of the droplet due to the divergence in the pressure at the Hagedorn temperature. The break up of a body involves \lq negative\rq\ pressures, which, like negative temperatures \cite{LL},  lie \lq above infinity\rq.
 \par The free energy per particle vanishes at the limiting temperature, while the energy and entropy per particle remain finite. The latter explains the lack of discontinuity in the density of states.\par However, all $N$-particle counterparts diverge at the limiting temperature. Whereas the $N$-particle entropy, 
\[\Sn(\beta,\Vn)=\Vn\left(\betah\left(\beta-\betah\right)^{-q/r}+\frac{q}{q-r}\left(\beta-\betah\right)^{-1/s}\right),\] diverges as $\left(\beta-\betah\right)^{-q/r}$, the same divergence as the energy (\ref{eq:EE}) and particle number (\ref{eq:N}), the $N$-particle entropy,
\[\Sn(\beta,\Vn)=\Vn\left(\betah\left(\beta-\betah\right)^{-1}+1-\ln\left(\beta-\betah\right)\right),\] diverges  more slowly as $\left(\beta-\betah\right)^{-1}$.\par  In the  former case, the square of the relative fluctuation tends to zero as $\left(\beta-\betah\right)^{1/s}$, while, in the latter case, it remains constant, as the Hagedorn temperature is approached. Therefore, in the latter case, the fluctuations in energy become as large as the energy itself. Nevertheless, in both cases the dispersion in energies diverge, while the dispersion in inverse temperature  both tend to zero as the limiting temperature is approached. This is a consequence of the thermodynamic uncertainty relation between the conjugate variables, energy and inverse temperature \cite{Lav91}. The temperature $T(E)$ of a system of energy $E$ becomes \lq\lq better and better defined---and equal to [$\Th$]---when $E$ grows larger and larger\rq\rq\cite{Hag}. In this sense, the Hagedorn thermostat is a perfect one \cite{Mor}.\par
Hence, the density of states,
\begin{equation}\Omega(E)=\exp\left[\frac{E}{q-r}\left(\frac{q}{T}-\frac{r}{\Th}\right)\right]\rightarrow\exp\left(\frac{E}{\Th}\right),\label{eq:Omega}
\end{equation}
approaches the Hagedorn spectrum asymptotically as $T\rightarrow\Th$ for $q\neq r$, while for $q=r$,
\[
\Omega(E)=\frac{E^V}{\Gamma(V+1)}e^{E/\Th}, 
\]
 contains a prefactor for the \lq normal\rq\ growth of the density of states, and  $V^Ve^{-V}$ has been replaced by the gamma function. \par
 In the non-asymptotic region of masses around $1$GeV \cite{Bron-bis}, (\ref{eq:Omega}) gives an effective temperature
 \[\frac{1}{T_{\mbox{\tiny{eff}}}}=\frac{1}{q-r}\left(\frac{q}{T}-\frac{r}{\Th}
 \right),\]
 with $rT/q<\Th$. As $T\rightarrow\Th$, $T_{\mbox{\tiny{eff}}}\rightarrow\Th$, which can be derived by imposing the asymptotic condition (\ref{eq:Coc}) on $E=VT^{q/r}$. Thus, the asymptotic, high energy, limit is characterized by the adiabats $V_0\Th^{1/s}=E/\Th=1/s$, where $V_0$ is the characteristic volume per particle of strong interactions.\par A pion gas with $s=\mbox{\small{$\frac{1}{3}$}}$, and a particle volume whose radius is the range of strong interactions, $1$ fm, has a limiting temperature, $\Th=180$ MeV. In the neighborhood of the \lq boiling point of hadronic\rq\ matter, particle creation becomes so violent that the temperature cannot increase further than $\Th$, no matter how much energy is pumped in  \cite{Hag-bis}. \par
 On the basis of the originally proposed hadron mass spectrum \cite{Hag,Fra}, \begin{equation}
 \Omega(E)=E^{\alpha-3}e^{\betah E},\label{eq:dos}
 \end{equation} with $\alpha<2$, it was argued that the system undergoes a second-order phase transition with the liberation of quarks \cite{Cab}. However, the partition function,
\[Z(\beta)=\left(\beta-\betah\right)^{-(\alpha-2)}=\int_0^{\infty}\,\frac{E^{\alpha-3}}{\Gamma(\alpha-2)}e^{-(\beta-\betah)E}\,dE,\]
 will only exist for values $\alpha>2$. This is also demanded by the property that the partition function be completely monotone so as to insure the positivity of the energy and the heat capacity.\par Consequently, the prefactor in  (\ref{eq:dos}) cannot be associated with a density of states, and there is no phase transition since $\Th$ is a limiting temperature, and not a critical temperature. In other words, \emph{the divergence of the thermodynamic potentials cannot be reconciled with a discontinuity because the temperature does not exist above\/} $\Th$.  The limiting temperature, $\Th$, is not an actual temperature because it is no longer a function of the energy, and, hence, cannot be estimated in terms of it. Consequently, the fundamental property of concavity of the entropy has been waived, and, so too, thermodynamics.\par

\end{document}